\documentclass[journal]{IEEEtran}

\usepackage{cite}

\ifCLASSINFOpdf

\else
\usepackage[dvips]{graphicx} 
\fi

\usepackage[cmex10]{amsmath}
\usepackage{amsthm,amssymb}
\usepackage{amsfonts}
\usepackage{dsfont}
\DeclareMathOperator*{\argmax}{arg\,max}

\begin{document}
\title{On Eavesdropper-Tolerance Capability of Two-Hop Wireless Networks}

\author{Yuanyu~Zhang,
        Yulong~Shen, Hua~Wang 
        and~Xiaohong~Jiang,~\IEEEmembership{Senior~Memeber,~IEEE}% <-this % stops a space
\thanks{Y.~Zhang and Y.~Shen are with the School of Computer Science and Technology,
        Xidian University, China. E-mail:yy90zhang@gmail.com;
        ylshen@mail.xidian.edu.cn}% <-this % stops a space
\thanks{H.~Wang is with the Department of Maths and Computing, University of Southern Queensland,
        Australia. Email: Hua.Wang@usq.edu.au}
\thanks{X.~Jiang is with the School of Systems Information Science,
        Future University Hakodate, 116-2, Kameda Nakano-Cho,
        Hakodate, Hokkaido, 041-8655, Japan. E-mail:jiang@fun.ac.jp.}% <-this % stops a space
}

\maketitle

\begin{abstract}
Two-hop  wireless network serves as the basic network model for the study of general wireless networks, 
while cooperative jamming is a promising scheme to achieve the physical layer security. 
This paper establishes a theoretical framework for the study of eavesdropper-tolerance capability 
(i.e., the exact maximum number of eavesdroppers that can be tolerated) in a two-hop wireless network, 
where the cooperative jamming is adopted to ensure security defined by secrecy outage probability (SOP) 
and opportunistic relaying is adopted to guarantee reliability defined by transmission outage probability (TOP). 
For the concerned network, closed form modeling for both SOP and TOP is first conducted based on the 
Central Limit Theorem. With the help of SOP and TOP models and also the Stochastic Ordering Theory, 
the model for eavesdropper-tolerance capability analysis is then developed. Finally, extensive simulation 
and numerical results are provided to illustrate the efficiency of our theoretical framework as well as 
the eavesdropper-tolerance capability of the concerned network from adopting cooperative jamming and 
opportunistic relaying.

\end{abstract}

% \fi
\IEEEpeerreviewmaketitle

\section{Introduction} \label{sec_1}
\IEEEPARstart{T}{wo-hop} wireless networks, in which a source can communicate with its destination directly or 
via a intermediate relay, have been a class of basic and attractive network scenarios \cite{narayanan2006two}. 
More importantly, the performance analysis in such two-hop networks lays the groundwork for the study in general 
multi-hop wireless networks. Due to the broadcast nature of wireless channels and the increasing demand for 
exchanging confidential information, ensuring secure and reliable transmission in such wireless networks has 
become a challenging yet critical task in practice, especially for those applications demanding high security 
and reliability, such as battle command, emergency treatment and disaster relief.

Traditionally, information is secured  above the physical layer by applying cryptography\cite{Stallings2010} or 
other approaches \cite{Xie2013}. The idea of cryptography is to encrypt the information through a cryptographic algorithm (e.g., RSA and AES) that is hard to break in 
practice by any eavesdropper with limited computing power and without the secret key. These 
schemes are therefore termed \textit{computationally secure}\cite{Diffie1976}, since they are built around the 
\textit{unproven} computational hardness assumption. However, recent advances in computing power (e.g., quantum computing) 
could make it possible to break such difficult cryptographic algorithms \cite{Shor1994} and thus the demand for 
everlasting security in modern wireless communications becomes more and more urgent. That is why there is an 
increasing interest recently in physical layer security, behind which the fundamental idea is to exploit 
the inherent physical characteristics of communication channels to provide \textit{information-theoretic} security to the 
legitimate transmissions without the assistance of a secret key \cite{Bloch2008,Mukherjee2010}. It is more important 
that no limitations are assumed for the eavesdroppers in terms of the computing power or network parameter knowledge. 
Moreover, the physical layer security approaches can offer some significant advantages over the traditional cryptographic 
scheme, like no need to employ complicated cryptographic algorithms and guaranteeing an everlasting security without 
applying key distribution and management, which is extremely expensive and difficult for large scale decentralized 
networks. Additionally, physical layer techniques can be used with cryptographic approaches in a complementary way 
and thus can augment the security achieved by cryptography. Therefore, physical layer approaches have been very 
promising in guaranteeing a strong form of security in wireless communications.
 
In the seminal work \cite{Wyner1975} on the physical layer security, Wyner introduced the wire-tap channel model
where the source transmits messages to the intended receiver over a discrete memoryless main channel which is 
wire-tapped by an eavesdropper (wiretapper) through another discrete memoryless channel, called wiretap channel. 
This work was later generalized to the broadcast model in \cite {Csiszar1978} and to the Gaussian setting in \cite{Leung1978}.  
These works indicated that perfect secrecy can be achieved if the intended receiver has a better channel than 
the eavesdropper, which however can hardly be satisfied in practice. Thus, many works sought to explore the possibility 
of secure transmission when the eavesdropper observes a better channel. Maurer \cite{Maurer1993} showed that perfect 
secrecy is achievable when the eavesdropper enjoys a better channel by generating a secret key over a public and 
error-free feedback channel. Nevertheless, this work is treated as a further step in the direction of public-key 
cryptology. Hero \cite{Hero2003} introduced the space–time coding over multiple antennas for secure communication 
and artificial noise injection strategy was first proposed by Negi and Goel \cite{Negi2005,Goel2005}, where the 
noise generated by the extra antennas of the transmitter such that only the eavesdropper channel is degraded. 
However, due to the cost of deploying multiple antennas and designing efficient noise, these schemes are not 
suitable for large scale wireless network with nodes of single antenna. Barros and Rodrigues \textit{et al.}\cite{Barros2006} 
analyzed the secrecy outage probability and outage secrecy capacity of a quasi-static Rayleigh fading channel and 
showed that fading alone can guarantee the information-theoretic security even when the eavesdropper has a better 
average SNR than the legitimate receiver. Tekin and Yener \cite{Tekin08} introduced the \textit{cooperative jamming} 
scheme where a nontransmitting user can increase the secrecy capacity by transmitting jamming signal instead of its 
codewords to confuse the eavesdropper. Since random noise can be generated by helper nodes rather than extra antennas, 
cooperating jamming has been widely introduced to enhance the physical layer security in wireless networks \cite{Zhou2011,
Capar2012,Dong2010,Li2011,BHan2013,Vilela2011,2Vilela2011,Huang2011,Krikidis2009,Ding2011,Bassily2013}.

By now, various works have been dedicated to explore the security performances in wireless networks with cooperative jamming. 
For instance, the per-node secure throughput in large decentralized networks was explored in \cite{Vasudevan2010,Zhou2011,Capar2012}, 
the secrecy capacity maximization problem was investigated in \cite{Dong2010,Li2011,BHan2013} based on cooperative communication, 
how to design efficient jamming strategies in terms of power or position of jamming was analyzed in \cite{Vilela2011,2Vilela2011,Huang2011}, 
the opportunistic selection and use of the relays to enhance the physical layer security was studied in\cite{Krikidis2009,Ding2011,Bassily2013}.
However, to the best of our knowledge, relatively fewer works consider the performance limits of the eavesdropper-tolerance
capability of a network. As shown in \cite{Haenggi2008,Liang2011}, the density of the eavesdroppers has a dramatic 
impact on the connectivity of secrecy graph and the secrecy throughput, which implies that the number of eavesdroppers 
present in the network is critical in guaranteeing the network security. Knowing the relationship between the eavesdropper-tolerance 
capability and other network parameters not only plays an important role in the security performance analysis of the 
network but also serve as the guideline on determining the system parameters to build a secure network for the designers. 
Therefore, we focus on the eavesdropper-tolerance capability study of a two-hop wireless network in this paper.

The related works regarding eavesdropper-tolerance capability can be classified into two categories according to 
the network size. For infinite network scenarios, the scaling law of eavesdropper-tolerance capability against the 
per-node throughput was studied in \cite{Vasudevan2010} by constructing a highway system. By cooperative jamming,
Goeckel \textit{et al.} \cite{Goeckel2011} considered one source-destination pair with opportunistic relaying scheme\cite{Bletsas2006}, 
where the best relay is selected among the available relays based on some policy in terms of their channels to the 
source and destination, and analyzed the asymptotic behavior of eavesdropper-tolerance capability
as the number of relays goes to infinity. However, the metrics used in their paper cannot fully reflect the security 
and reliability of the end-to-end transmission. This work was later generalized to a scenario with multiple source-destination 
pairs where artificial noises are generated from concurrent transmitters \cite{Shei2012}. For finite network scenarios, 
Shen \textit{et al.}\cite{Shen2013} proposed a flexible relay selection scheme and derived the \textit{lower bound} on the eavesdropper-tolerance 
capability. However, it is notable that all the above works have focused on either the order-sense scaling law results 
for infinite networks, or bounds for finite networks. Such order sense results or bounds are certainly important but 
cannot reflect the actual eavesdropper-tolerance capability in more practical network scenarios with finite nodes, 
which is more important for the system designers. In our previous work \cite{Zhang2013}, we considered a random relay 
selection scheme and derived the \textit{exact} eavesdropper-tolerance capability, which can exactly tell us how many 
eavesdroppers a network can tolerate at most for a desired level of security and reliability. However, the results showed 
that with the random relay selection, the eavesdropper-tolerance performance is not good, especially for small-scale 
networks and high security/reliability requirement.  

In this paper, we establish a theoretical framework to explore the  eavesdropper-tolerance capability in 
a two-hop wireless network, where the cooperative jamming is adopted to ensure security defined by secrecy outage probability (SOP)
and opportunistic relaying is adopted to guarantee reliability defined by transmission outage probability (TOP). Different 
from \cite{Goeckel2011}, we use different outage probability metrics that can fully characterize the security and reliability 
of the end-to-end transmission. More importantly, we consider the inherent channel dependence of the transmissions in two hops, 
which is critical in determining the exact eavesdropper-tolerance capability. Our contributions can be summarized as follows:
\begin{itemize}
\item We first apply the the Central Limit Theorem to develop the closed form models for both SOP and TOP of a source-destination transmission.
\item Based on the SOP and TOP models and also the Stochastic Ordering Theory, we then conduct analysis to reveal the monotonicity properties of 
SOP and TOP. With the help of such properties, the model for eavesdropper-tolerance capability is derived.
\item A simulator is developed to validate the efficiency of our theoretical framework and numerical results are also provided to 
illustrate the eavesdropper-tolerance capability of the concerned network from adopting cooperative jamming and opportunistic relaying. 
\end{itemize}

The reminder of the paper is organized as follows. Section \ref{sec_2} introduces the system model and problem formulation.
In Section \ref{sec_3}, we conduct the closed form modeling of SOP and TOP of the end-to-end transmission. The model for
eavesdropper-tolerance capability analysis is developed in Section \ref{sec_4}. Section \ref{sec_5} presents the simulation 
and numerical results to validate our theoretical model and Section \ref{sec_6} concludes the paper.

\section{System Model and Problem Formulation} \label{sec_2}
\subsection{System Model and Assumptions}\label{sec_2_1}
\begin{figure}[!t]
\centering
\includegraphics[width=3 in]{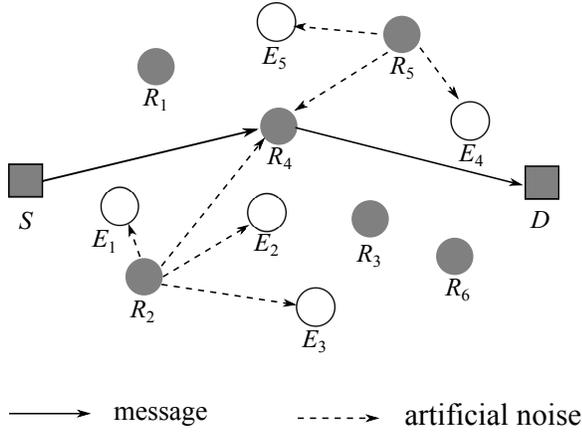}
\caption{System scenario: Source $S$ is transmitting message to the destination
$D$ with the help of relays $R_1,R_2,\cdots,R_n $ ($n=6$ in this figure) while
eavesdroppers $E_1,E_2,\cdots,E_m$ ($m=5$ in this figure) are attempting to
intercept the message. In this figure, $R_4$ is the message relay and $R_2$,
$R_5$ are noise-generating relays.}
\label{fig_model}
\end{figure}

As depicted in Fig.\ref{fig_model}, we consider a two-hop wireless network scenario consisting 
of a source node \emph{S}, a destination node \emph{D}, $n$ legitimate half-duplex relays $R_1,R_2,
\cdots,R_n$ that cannot transmit and receive at the same time and $m$ passive and independently-operating 
eavesdroppers $E_1,E_2,\cdots,E_m$. We assume that the direct link between \emph{S} and \emph{D} 
does not exist due to the deep fading and thus \emph{S} needs to transmit messages to \emph{D} via one of 
the relays.  Each of the eavesdroppers attempts to intercept the messages on its own. Meanwhile, 
some of the remaining $n-1$ relays will be selected to generate artificial noise to suppress the 
eavesdroppers during the transmission. We aim to ensure both the secure and reliable transmission 
from \emph{S} to \emph{D} against these eavesdroppers of unknown channel and location information. 

A slow and flat block Rayleigh Fading environment is assumed, where the channel remains static for
one coherence interval and varies randomly and independently from interval to interval. Thus, the
channel from a transmitter \emph{A} to a receiver \emph{B} can be represented by a complex zero-mean
Gaussian random variable $h_{A,B}$ and the corresponding channel gain $|h_{A,B}|^2$ is an exponential
random variable. Without loss of generality, we assume that $|h_{A,B}|^2=|h_{B,A}|^2$ and 
$\mathbb E\big[|h_{A,B}|^2\big]=1$, where $\mathbb E\big[\cdot\big]$ stands for the expectation 
operator. It is assumed that the source \emph{S} and the relays transmit with the same power $P_t$. 
In addition, we assume the network is interference-limited and thus the noise at each receiver
is negligible. Therefore, when \emph{A} is transmitting and relays with indices in $\mathcal R$ are 
generating noise, the received signal-to-interference ratio (\emph{SIR}) at a receiver \emph{B} 
can be formulated as
\begin{align*}
SIR_{A,B}=\frac{P_t\cdot|h_{A,B}|^2}{\sum_{j\in\mathcal R} P_t\cdot |h_{R_j,B}|^2}=\frac{|h_{A,B}|^2}{\sum_{j\in\mathcal R} |h_{R_j,B}|^2}.
\end{align*}  

For the eavesdroppers and legitimate receivers, we use positive $\gamma_e$ and $\gamma$ respectively 
to denote the minimum \textit{SIR} required to recover the received message. That is, a legitimate
receiver (eavesdropper) is able to decode the transmitted message if and only if its received \textit{SIR} 
exceeds $\gamma$ ($\gamma_e$). This \textit{SIR} threshold scheme can be easily mapped to
the Wyner's encoding scheme where the transmitter chooses two rates, the rate of transmitted codewords 
$R_t$ and the rate of the confidential message $R_s$ \cite{Wyner1975},\cite{Zhou2011}. The rate difference
$R_e = R_t - R_s$ reflects the cost of securing the message against the eavesdroppers. The conversions 
between the thresholds and the code rates are as follows:
\begin{align*}
\gamma = 2^{R_t}-1,\\
\gamma_e = 2^{R_e}-1.
\end{align*}
Therefore, the results in this paper also applies to the Wyner's encoding scheme.

In order to improve the link condition from \emph{S} to \emph{D},  an opportunistic relaying 
scheme is adopted, where the best relay $R_b$ is selected by a timer-based method explained in \cite{Bletsas2006} 
to forward messages. Here, $b$ is given by  
\begin{align*}
b\overset{\underset{\Delta}{}}{=}\argmax_j~ min\{|h_{S,R_j}|^2,|h_{R_j,D}|^2\}.
\end{align*}
The transmission then can be conducted in two phases. In the first phase, \emph{S} transmits the message
to $R_b$. At the same time, relays with indices in $\mathcal R_1=\big\{j\big|j\neq b,|h_{R_j,R_b}|^2<\tau\big\}$,
where $\tau$ is the noise-generating threshold to control the interference at legitimate receivers, 
generate artificial noise to suppress the eavesdroppers. Analogous to the first phase, $R_b$ forwards 
its received message to $D$ with relays whose indices are in  $\mathcal R_2=\big\{j\big|j\neq b,
|h_{R_j,D}|^2<\tau\big\}$ generating noise to assist the transmission in the second phase.

\subsection{Problem Formulation} \label{sec_2_2}

In this subsection, we first introduce the concepts of TOP and SOP of the concerned network, based on which 
we then formulate our problem regarding the eavesdropper-tolerance capability in this paper. 

To fully characterize the security and reliability performances of the transmission, we adopt the same outage 
definitions in \cite{Zhou2011}. Consider the direct link from a transmitter \emph{A} to a legitimate receiver \emph{B}. 
We say transmission outage happens if \emph{B} cannot decode the message (i.e., $SIR_{A,B}<\gamma$) and secrecy outage 
happens if at least one of the eavesdroppers (say $E_i$) can decode the message (i.e., $SIR_{A,E_i}\geq \gamma_e$).
It is shown in \cite{Koyluoglu2012} that securing each of the individual links is sufficient to secure the 
end-to-end path. Thus, the secrecy (transmission) outage of the $S\rightarrow R_b\rightarrow D$ link occurs if either 
$S\rightarrow R_b$ or $R_b \rightarrow D$ suffers from secrecy (transmission) outage. Then we can introduce the 
following definitions:

\begin{itemize}
\item \textbf{TOP for opportunistic relaying $P_{bst}^{to}$}:  This probability 
is defined as the probability that the transmission outage of the $S\rightarrow R_b\rightarrow D$ link happens under
the opportunistic relaying scheme.
\item \textbf{SOP for opportunistic relaying $P_{bst}^{so}$}: This probability is defined 
as the probability that the secrecy outage of the $S\rightarrow R_b\rightarrow D$ link happens under the opportunistic 
relaying scheme.
\end{itemize}

Based on the above definitions, $P_{bst}^{to}$ and $P_{bst}^{so}$ can be formulated as 

\begin{IEEEeqnarray}{rCl}
P_{bst}^{to}&=& \mathbb P\left( SIR_{S,R_b}<\gamma\cup SIR_{R_b,D}<\gamma\right) \label{eq_df_ToutPro}\\
&=&1-\mathbb P\left( SIR_{S,R_b}\geq \gamma,SIR_{R_b,D}\geq \gamma\right)\nonumber\\
&=&1-\mathbb P\left( \frac{|h_{S,R_b}|^2}{\sum_{j\in \mathcal R_1}|h_{R_j,R_b}|^2}\geq \gamma,\frac{|h_{R_b,D}|^2}{\sum_{j\in \mathcal R_2}|h_{R_j,D}|^2}\geq \gamma\right)\nonumber
\end{IEEEeqnarray}
and
\begin{IEEEeqnarray}{rCl}
P_{bst}^{so}&=& \mathbb P\left( \bigcup_{i=1}^{m} \left\{SIR_{S,E_i}\geq \gamma_e\right\}\cup \bigcup_{i=1}^{m}\left\{SIR_{R_b,E_i}\geq\gamma_e\right\}\right) \label{eq_df_SoutPro}\\
&=&1-\mathbb P\left( \bigcap_{i=1}^{m}\left\{SIR_{S,E_i}<\gamma_e\right\},\bigcap_{i=1}^{m}\left\{SIR_{R_b,E_i}<\gamma_e\right\}\right)\nonumber\\
&\overset{(a)}=&1-\left[\mathbb P\Big( \bigcap_{i=1}^{m}\left\{SIR_{S,E_i}<\gamma_e\right\}\Big)\right]^2\nonumber\\
&=&1-\left[\mathbb P\Big( \bigcap_{i=1}^{m}\Big\{\frac{|h_{S,E_i}|^2}{\sum_{j\in\mathcal R_1} |h_{R_j,E_i}|^2}<\gamma_e\Big\}\Big)\right]^2\nonumber
\end{IEEEeqnarray}
where $\mathbb P(\cdot)$ stands for the probability operator and $(a)$ follows since the received power of each 
eavesdropper in two phases are independent and identically distributed. It is notable that the second $\mathbb P(\cdot)$ 
term in (\ref{eq_df_ToutPro}) cannot be formulated as 
\begin{align*}
\mathbb P\left( SIR_{S,R_b}\geq \gamma\right)\mathbb P\left(SIR_{R_b,D}\geq \gamma\right),
\end{align*}
since $SIR_{S,R_b}$ and $SIR_{R_b,D}$ are dependent, as will be observed in Appendix A.

Since security and reliability are two important metrics in network design, we use the SOP 
constraint $\varepsilon_s$ and TOP constraint $\varepsilon_t$ to represent the 
security and reliability requirements of the end-to-end transmission. We say that the transmission from \emph{S} to \emph{D} 
is \emph{secure} \emph{if and only if} $P_{bst}^{so} \leq \varepsilon_s$ and it is \emph{reliable} \emph{if and only if} 
$P_{bst}^{to} \leq \varepsilon_t$. Notice that larger  $\varepsilon_s$ and $\varepsilon_t$  represent 
less stringent security and reliability requirements. In this paper, we aim to determine the \emph{exact} 
$P_{bst}^{to}$ and $P_{bst}^{so}$, which can be then used 
to determine the \emph{exact} eavesdropper-tolerance capability while ensuring both the \emph{reliable} and \emph{secure} 
end-to-end transmission. We use ${m}^*_{bst}$ to represent the eavesdropper-tolerance capability for the opportunistic 
relaying scheme hereafter.

Based on the above observations, we are now ready to formulate our problem. From the definition of $P_{bst}^{to}$
and $P_{bst}^{so}$, we can see that when given the number of system relays $n$, the \emph{SIR} thresholds $\gamma$, 
$\gamma_e$, the security requirement $\varepsilon_s$ and reliability requirement $\varepsilon_t$,  ${m}^*_{bst}$ 
only depends on the noise-generating threshold $\tau$. Thus, we define the maximum number of eavesdroppers that can 
be tolerated \emph{for a specified} $\tau$ by 
\begin{align*}
M_{bst}(\tau)=max\{m: P_{bst}^{so}(n,m,\tau)\leq \varepsilon_s\}.
\end{align*} 
Now the considered problem can be formulated as 
\begin{equation}
\begin{aligned}
& \underset{\tau}{\text{maximize}}
& & M_{bst}(\tau)  \\
& \text{subject to}
& & P_{bst}^{to}(n,\tau)\leq \varepsilon_t,\tau \geq 0\\
&
& &\varepsilon_t \in [0,1], \varepsilon_s \in [0,1]
\end{aligned}
\label{eq_profor}
\end{equation}
where $P_{bst}^{to}$ and $P_{bst}^{so}$ are regarded as functions. That is, we want to maximize $M_{bst}(\tau)$
over $\tau$. We use $\tau_{bst}^b$ to represent the optimal $\tau$ that maximizes $M_{bst}(\tau)$ for opportunistic 
relaying scheme and thus we have ${m}^*_{bst}=M_{bst}(\tau_{bst}^b)$.

In order to explore the efficiency of the opportunistic relaying scheme, 
we also give the eavesdropper-tolerance capability of the same network scenario but with a random relay selection 
scheme as a comparison, which is considered in \cite{Zhang2013}. Similarly, for random relay selection scheme, we define the TOP
by $P_{ran}^{to}$, the SOP by $P_{ran}^{so}$, the optimal $\tau$ by $\tau_{ran}^{b}$ and the eavesdropper-tolerance 
capability by ${m}^*_{ran}$.

\section{Outage Performances} \label{sec_3}
In this section we determine the TOP $P_{bst}^{to}$ and SOP $P_{bst}^{so}$ of the network with opportunistic relaying 
scheme based on some theoretical analysis. Applying the same approach, we also give the outage probabilities of the network 
with random relay selection scheme.

\subsection{SOP and TOP For Opportunistic Relaying}\label{sec_3_1}
Before determining the TOP of a network with opportunistic relaying, 
we first define the total interference at the legitimate receiver in two phases by 
\begin{align*}
I_1=\sum_{j\in \mathcal R_1}|h_{R_j,R_b}|^2,I_2=\sum_{j\in \mathcal R_2}|h_{R_j,D}|^2.
\end{align*} 
Then, we establish the following lemmas regarding the probability distribution of $I_1$, $I_2$ and an 
important joint probability of the channel gains in two phases, which is critical in determining $P_{bst}^{to}$.

\emph{Lemma} 1: For one message transmission from  \emph{S} to \emph{D}, the total 
interference $I_1$ and $I_2$ are independent and identically distributed, and can be approximated 
by a normal random variable. Thus, the corresponding pdf is given by
\begin{align*}
f(x)\approx \hat{f}(x)=\frac{e^{\frac{(x-\mu)^2}{2\sigma^2}}}{\sigma\sqrt{2\pi}},
\end{align*}
where 
\begin{align*}
\mu=(n-1)\Big[1-(1+\tau)e^{-\tau}\Big]
\end{align*}
is the mean and 
\begin{align*}
\sigma=\sqrt{(n-1)\Big[1-\tau^{2}e^{-\tau}-(1+\tau)^{2}e^{-2\tau}\Big]}
\end{align*}
is the standard derivation of the normal random variable.

\emph{Lemma} 2: For one message transmission from \emph{S} to \emph{D}, the joint probability that $|h_{S,R_b}|^2$ is greater 
than some constant $x \geq 0$ and  $|h_{R_b,D}|^2$ is greater than some constant $y \geq 0$ can be determined as
\begin{IEEEeqnarray}{rCl}
\IEEEeqnarraymulticol{3}{l}{\mathbb P\Big(|h_{S,R_b}|^2 \geq x ,|h_{R_b,D}|^2\geq y\Big)} \nonumber\\
&=& 1-(1-e^{-2max\{x,y\}})^{n}\nonumber \\
&&+ne^{-max\{x,y\}}\Big[\varphi(n,min\{x,y\})-\varphi(n,max\{x,y\})\Big],\nonumber
\end{IEEEeqnarray}
where
\begin{align*}
\varphi(n,x)= e^{-x}{}_2F_1\left(\frac{1}{2},1-n;\frac{3}{2};e^{-2x}\right)
\end{align*}
and ${}_2F_1$ is the Gaussian hypergeometric function.

\emph{Remark} 1: Since $S$ and relays transmit with the same power, $P_t$ can be reduced 
in determining the TOP as shown in (\ref{eq_df_ToutPro}), and thus it is not considered in Lemma 1. 
The proofs of the above lemmas can be found in Appendix A.

For a two-hop wireless network with opportunistic relaying scheme, we are now ready to derive its TOP $P_{bst}^{to}$ and SOP $P_{bst}^{so}$ of the end-to-end transmission 
based on Lemma 1 and Lemma 2.

\textbf{Theorem 1}: Consider the network scenario in Fig.\ref{fig_model} with opportunistic relaying
scheme. The TOP $P_{bst}^{to}$ and SOP $P_{bst}^{so}$ can be given by
\begin{IEEEeqnarray}{rCl}
&P_{bst}^{to}\approx& 2\int_{0}^{(n-1)\tau}g(n,\gamma,x)\hat{f}(x)\left[\Phi(\frac{x-\mu}{\sigma})-\Phi(-\frac{\mu}{\sigma})\right]dx \nonumber \\
&&-2\int_{0}^{(n-1)\tau}\int_{0}^{x}ne^{-\gamma x}\varphi(n,\gamma y)\hat{f}(x)\hat{f}(y)dydx
\label{eq_bst_to1}
\end{IEEEeqnarray}
and
\begin{IEEEeqnarray}{rCl}
P_{bst}^{so}=1-\left(\sum_{k=1}^{m}\binom{m}{k}(-1)^k\Big[(1-e^{-\tau})c^{k}+e^{-\tau}\Big]^{n-1}\right)^2
\label{eq_bst_so1}
\end{IEEEeqnarray}
where
\begin{align*}
c=\frac{1}{1+\gamma_e},~\hat{f}(x)=\frac{e^{\frac{(x-\mu)^2}{2\sigma^2}}}{\sigma\sqrt{2\pi}},
\end{align*}
\begin{align*}
\Phi(x)=\frac{1}{\sqrt{2\pi}}\int_{-\infty}^{x}e^{-\frac{t^2}{2}}dt,
\end{align*}
\begin{align*}
\mu=(n-1)\Big[1-(1+\tau)e^{-\tau}\Big],
\end{align*}
\begin{align*}
\sigma=\sqrt{(n-1)\Big[1-\tau^{2}e^{-\tau}-(1+\tau)^{2}e^{-2\tau}\Big]},
\end{align*}
\begin{align*}
g(n,\gamma,x)=(1-e^{-2\gamma x})^n+ne^{-\gamma x}\varphi(n,\gamma x),
\end{align*} 
\begin{align*}
\varphi(n,x)= e^{-x}{}_2F_1\left(\frac{1}{2},1-n;\frac{3}{2};e^{-2x}\right)
\end{align*}
and ${}_2F_1$ is the Gaussian hypergeometric function.

\begin{IEEEproof}
1) We first prove the  $P_{bst}^{to}$ in (\ref{eq_bst_to1}). According to the definition in (\ref{eq_df_ToutPro}),
we have 
\begin{IEEEeqnarray}{rCl}
P_{bst}^{to}&=& 1-\mathbb P\Big(SIR_{S,R_b}\geq \gamma,SIR_{R_b,D}\geq \gamma\Big) \nonumber \\
&=& 1-\mathbb P\left(|h_{S,R_b}|^2\geq \gamma I_1,|h_{R_b,D}|^2\geq \gamma I_2\right) \nonumber 
\end{IEEEeqnarray}
Applying the law of total probability, we have 
\begin{IEEEeqnarray}{rCl}
&P_{bst}^{to}&= 1-\mathbb E_{I_1,I_2}\Big[\mathbb P\left(|h_{S,R_b}|^2\geq \gamma I_1,|h_{R_b,D}|^2\geq \gamma I_2\right)\Big]\label{eq_bst_to2} \\
&\overset{(b)}\approx& 1-\int_{0}^{(n-1)\tau}\int_{0}^{(n-1)\tau} \mathbb P\Big(|h_{S,R_b}|^2\geq \gamma x,|h_{R_b,D}|^2\geq \gamma y\Big) \nonumber\\
&&\times\hat{f}(x)\hat{f}(y)dydx \nonumber \\
&\overset{(c)}=&2\int_{0}^{(n-1)\tau}\int_{0}^{x}\Bigg\{(1-e^{-2\gamma x})^n \nonumber \\
&&-ne^{-\gamma x}\Big[\varphi(n,\gamma y)-\varphi(n,\gamma x)\Big]\Bigg\}\hat{f}(x)\hat{f}(y)dydx\nonumber \\
&=&2\int_{0}^{(n-1)\tau}\int_{0}^{x}g(n,\gamma,x)\hat{f}(x)\hat{f}(y)dydx \nonumber\\
&&-2\int_{0}^{(n-1)\tau}\int_{0}^{x}ne^{-\gamma x}\varphi(n,\gamma y)\hat{f}(x)\hat{f}(y)dydx\nonumber \\
&=&2\int_{0}^{(n-1)\tau}g(n,\gamma,x)\hat{f}(x)\left[\Phi(\frac{x-\mu}{\sigma})-\Phi(-\frac{\mu}{\sigma})\right]dx \nonumber\\
&&-2\int_{0}^{(n-1)\tau}\int_{0}^{x}ne^{-\gamma x}\varphi(n,\gamma y)\hat{f}(x)\hat{f}(y)dydx \nonumber, 
\end{IEEEeqnarray}
where $(b)$ is due to Lemma~1 and $(c)$ follows after applying Lemma 2.

2) Now we proceed to prove the $P_{bst}^{so}$ in (\ref{eq_bst_so1}). According to the definition in 
(\ref{eq_df_SoutPro}), we first need to derive the probability $\mathbb P\Big( \bigcap_{i=1}^{m}\Big\{\frac{|h_{S,E_i}|^2}
{\sum_{j\in\mathcal R_1} |h_{R_j,E_i}|^2}<\gamma_e\Big\}\Big)$.

Note that the number of noise-generating relays in the first phase $|\mathcal R_1|$ follows the binomial distribution 
$B(n-1,1-e^{-\tau})$. Now, we define the event that there are $l$ noise-generating relays in the first phase 
(i.e., $|\mathcal R_1|=l$) by $B_l$ and thus we have
\begin{IEEEeqnarray}{rCl}
&&\mathbb P\Big( \bigcap_{i=1}^{m}\Big\{\frac{|h_{S,E_i}|^2}{\sum_{j\in\mathcal R_1} |h_{R_j,E_i}|^2}<\gamma_e\Big\}\Big) \label{eq_SoutPro_pf}\\
&=&\sum_{l=0}^{n-1}\mathbb P\Big( \bigcap_{i=1}^{m}\Big\{\frac{|h_{S,E_i}|^2}{\sum_{j\in\mathcal R_1} |h_{R_j,E_i}|^2}<\gamma_e\Big\}\Big|B_l\Big)\mathbb P(B_l)\nonumber\\
&\overset{(d)}=&\sum_{l=0}^{n-1}\prod_{i=1}^{m}\mathbb P\Big(\frac{|h_{S,E_i}|^2}{\sum_{j\in\mathcal R_1} |h_{R_j,E_i}|^2}<\gamma_e\Big|B_l\Big)\mathbb P(B_l)\nonumber\\
&\overset{(e)}=&\sum_{l=0}^{n-1}\prod_{i=1}^{m}\mathbb E\left[1-e^{-\gamma_e\sum_{j\in\mathcal R_1} |h_{R_j,E_i}|^2}\right]\mathbb P(B_l)\nonumber\\
&\overset{(f)}=&\sum_{l=0}^{n-1}\prod_{i=1}^{m}\Bigg(1-\prod_{j\in\mathcal R_1}\mathbb E\left[e^{-\gamma_e |h_{R_j,E_i}|^2}\right]\Bigg)\mathbb P(B_l)\nonumber\\
&=&\sum_{l=0}^{n-1}\left[1-\left(\frac{1}{1+\gamma_e}\right)^l\right]^m\binom{n-1}{l}(1-e^{-\tau})^l(e^{-\tau})^{n-1-l}\nonumber\\
&=&\sum_{l=0}^{n-1}\sum_{k=1}^{m}\binom{m}{k}(-1)^k\binom{n-1}{l}c^{lk}(1-e^{-\tau})^l(e^{-\tau})^{n-1-l}\nonumber\\
&\overset{(g)}=&\sum_{k=1}^{m}\binom{m}{k}(-1)^k\Big[(1-e^{-\tau})c^{k}+e^{-\tau}\Big]^{n-1}\nonumber
\end{IEEEeqnarray} 
where $(d)$ follows since all the $\{SIR_{S,E_i},i=1,\cdots,m\}$ are conditionally independent given event $B_l$, 
$(e)$ follows by applying the law of total probability and the expectation is computed with respect to $\{|h_{R_j,E_i}|^2,j\in\mathcal R_1\}$, 
$(f)$ follows since all the $|h_{R_j,E_i}|^2$ are independent and identically distributed and $(g)$ follows by applying the binomial theorem. 
Therefore, (\ref{eq_bst_so1}) follows after substituting (\ref{eq_SoutPro_pf}) into (\ref{eq_df_SoutPro}).   
\end{IEEEproof}

\subsection{SOP and TOP For Random Relay Selection} \label{sec_3_2}

Applying the same approach, we now can establish the following lemma about the TOP and SOP under the 
random relay selection scheme.

\emph{Lemma} 3: Consider the network scenario in Fig.\ref{fig_model} with random relay selection scheme. The TOP
 $P_{ran}^{to}$ and SOP $P_{ran}^{so}$ can be given by 
\begin{IEEEeqnarray}{rCl}
&P_{ran}^{to}=& 1-\left(e^{-\tau}+\frac{1-e^{-(1+\gamma)\tau}}{1+\gamma}\right)^{2n-2}
\label{eq_ran_to}
\end{IEEEeqnarray}
and
\begin{IEEEeqnarray}{rCl}
P_{ran}^{so}=1-\left(\sum_{k=1}^{m}\binom{m}{k}(-1)^k\Big[(1-e^{-\tau})c^{k}+e^{-\tau}\Big]^{n-1}\right)^2
\label{eq_ran_so}
\end{IEEEeqnarray}
where $c=\frac{1}{1+\gamma_e}$.

\emph{Remark} 2: The distributions of $I_1$ and $I_2$ are not used in determining the $P_{ran}^{to}$ in (\ref{eq_ran_to}),
because the channel gains in two hops are independent. Therefore, we can give an exact TOP. 
The detailed proof can be found in \cite{Zhang2013}. It is also noticed that the SOP $P_{ran}^{so}$ 
in (\ref{eq_ran_so}) is identical to $P_{bst}^{so}$ in (\ref{eq_bst_so1}). This is because that the noise-generating schemes are 
identical in these two schemes and the message relay selection has no impact on the intercepting behavior of the eavesdroppers.

\section{Eavesdropper-Tolerance Capability}\label{sec_4}

Eavesdropper-tolerance capability characterizes how many eavesdroppers that can be tolerated at most by a wireless network with $n$
relays in order to guarantee the desired security requirement $\varepsilon_s$ and reliability requirement $\varepsilon_t$. In 
this section, we determine the eavesdropper-tolerance capability for opportunistic relaying scheme based on the problem formulation 
in section \ref{sec_2_2}. The eavesdropper-tolerance capability for random relay selection scheme is also provided by applying 
the same approach.

\subsection{Eavesdropper-Tolerance Capability for Opportunistic Relaying}\label{sec_4_1}
It can be observed from the transmission scheme in section \ref{sec_2_1} and the problem formulation in section \ref{sec_2_2} that
the noise-generating threshold $\tau$ is a critical parameter in determining the eavesdropper-tolerance capability. Too large
$\tau$ will do harm to the end-to-end transmission, while too small $\tau$ is not enough to interfere the eavesdroppers. 
Therefore, finding a optimal $\tau$ is the key to solving our considered problem. Before solving the problem, we establish 
the following lemma based on the Stochastic Ordering in \cite{Shaked2010}.

\emph{Lemma} 4:  Let $\mathbf{X}$ and $\mathbf{Y}$ be two N-dimensional random vectors such that 
\begin{align*}
\mathbb P(\mathbf{X}\in U)\leq \mathbb P(\mathbf{Y}\in U)~~\text{for all upper sets} ~~U\in \mathbb R^N.
\end{align*}
Then $\mathbf{X}$ is said to be \textit{smaller than $\mathbf{Y}$ in the usual stochastic order} (denoted by $\mathbf{X}\leq_{st} \mathbf{Y}$).
And for all increasing function $\phi$, we always have $\mathbb E[\phi(\mathbf{X})]\leq \mathbb E[\phi(\mathbf{Y})]$. 

Based on the above lemma, we then establish the following lemmas in terms of the monotonicity of 
SOP and TOP with respect to $\tau$.

\emph{Lemma} 5: The TOP $P_{bst}^{to}$ for opportunistic relaying scheme increases as $\tau$ increases.
\begin{IEEEproof}
For any $0<\tau_1<\tau_2$, we use random vector $\mathbf{I}_1=(I_1^1,I_2^1)$  to represent the interferences in two phases 
when the noise-generating threshold is $\tau_1$ and $\mathbf{I}_2=(I_1^2,I_2^2)$ to represent those interferences for $\tau_2$. 
For any upper set $U=\left\{(I_1,I_2)\big |I_1\geq x\geq 0, I_2\geq y\geq 0\right\}$, we always have
\begin{align*}
\mathbb P(\mathbf{I}_1\in U)= \mathbb P(I_1^1\geq x)\mathbb P(I_2^1\geq y)
\end{align*}
and
\begin{align*}
\mathbb P(\mathbf{I}_2\in U)= \mathbb P(I_1^2\geq x)\mathbb P(I_2^2\geq y).
\end{align*}
It is easy to see that $\mathbb P(I_1^1\geq x)< \mathbb P(I_1^2\geq x)$  and $\mathbb P(I_2^1\geq y)<\mathbb P(I_2^2\geq y)$, since 
more interference can be generated as $\tau$ increases. Therefore, we have $\mathbb P(\mathbf{I}_1\in U)< \mathbb P(\mathbf{I}_2\in U)$
and then $\mathbf{I}_1\leq_{st}\mathbf{I}_2$ according to Lemma $4$. Define the term $\mathbb P\left(|h_{S,R_b}|^2\geq \gamma I_1,|h_{R_b,D}|^2\geq \gamma I_2\right)$ 
in (\ref{eq_bst_to2}) by $\Gamma(\mathbf{I})$ which  decreases as $\mathbf{I}$ increases, where $\mathbf{I}=(I_1,I_2)$.
Thus, we have $\mathbb E[\Gamma(\mathbf{I}_1)]>\mathbb E[\Gamma(\mathbf{I}_2)]$ according to Lemma $4$. That is, for any $0<\tau_1<\tau_2$,
we always have $P_{bst}^{to}(\tau_1)<P_{bst}^{to}(\tau_2)$, which indicates the TOP $P_{bst}^{to}$ increases with $\tau$.
\end{IEEEproof}
 
\emph{Lemma} 6: The SOP $P_{bst}^{so}$ for opportunistic relaying scheme decreases as $\tau$ increases, whereas
increases as $m$ increases.
\begin{IEEEproof}
Notice that the step following $(f)$ in (\ref{eq_SoutPro_pf}) can also be written as 
\begin{align*}
\mathbb E\left[1-\left(\frac{1}{1+\gamma_e}\right)^{|\mathcal R_1|}\right],
\end{align*}
where the expectation is computed with respect to $|\mathcal R_1|$. For any $0\leq\tau_1<\tau_2$, we use two random variables $|\mathcal R_1^1|$ 
and $|\mathcal R_1^2|$ to represent the number of noise-generating relays in the first phase, where 
\begin{align*}
|\mathcal R_1^1|\sim B(n-1,1-e^{-\tau_1})
\end{align*}
and 
\begin{align*}
|\mathcal R_1^2|\sim B(n-1,1-e^{-\tau_2}). 
\end{align*} 
It is shown in \cite{Klenke2010} that $|\mathcal R_1^1| \leq_{st} |\mathcal R_1^2|$. Applying Lemma $4$ again, we can see that 
\begin{align*}
\mathbb E\left[1-\left(\frac{1}{1+\gamma_e}\right)^{|\mathcal R_1^1|}\right]<\mathbb E\left[1-\left(\frac{1}{1+\gamma_e}\right)^{|\mathcal R_1^2|}\right]
\end{align*}
Therefore, the SOP $P_{bst}^{so}$ decreases as $\tau$ increases.

Next, we consider the step following $(f)$ in  (\ref{eq_SoutPro_pf}) again. It is easy to see that the term 
\begin{align*}
1-(\frac{1}{1+\gamma_e})^l \in [0,1).
\end{align*}
Thus, the term $\left[1-\left(\frac{1}{1+\gamma_e}\right)^l\right]^m$ decreases with $m$. Therefore, the SOP $P_{bst}^{so}$ increases as $m$ increases.
\end{IEEEproof}

Define step $(g)$ in (\ref{eq_SoutPro_pf}) by a function $G(m,n,\tau)$. Then we can derive the eavesdropper-tolerance capability 
${m}^*_{bst}$ for opportunistic relaying scheme based on Lemma $5$ and Lemma $6$.

\textbf{Theorem 2}: Consider the network scenario in Fig.\ref{fig_model} with opportunistic relaying scheme. The eavesdropper-tolerance
capability under the security constraint $\varepsilon_s$ and reliability constraint $\varepsilon_t$ is 
\begin{IEEEeqnarray}{rCl}
{m}^*_{bst}= max\{m: G(m,n,\tau_{bst}^b)\geq \sqrt{1-\varepsilon_s}\},
\label{eq_theo2}
\end{IEEEeqnarray}
where 
\begin{align*}
G(m,n,\tau_{bst}^b)=\sum_{k=1}^{m}\binom{m}{k}(-1)^k\Big[(1-e^{-\tau_{bst}^b})c^{k}+e^{-\tau_{bst}^b}\Big]^{n-1},
\end{align*}
$c=\frac{1}{1+\gamma_e}$ and $\tau_{bst}^b$ is the solution of $P_{bst}^{to}=\varepsilon_t$.

\begin{IEEEproof}
As shown in (\ref{eq_profor}), we need to find the optimal $\tau$ that maximizes $M_{bst}(\tau)$, where 
\begin{align*}
M_{bst}(\tau)=max\{m: G(m,n,\tau)\geq \sqrt{1-\varepsilon_s}\},
\end{align*}
according to its definition. Since the TOP $P_{bst}^{to}$ increases with $\tau$ according to Lemma $5$, 
in order to guarantee the reliability (i.e., $P_{bst}^{to}\leq \varepsilon_t$), $\tau$ must take values in the region $[0,\tau_m]$, 
where $\tau_m$ is the solution of $P_{bst}^{to}=\varepsilon_t$.

Next, we need to prove that $\tau_m$ is the optimal $\tau$ (i.e.,$\tau_{bst}^{b}=\tau_m$). That is, for any $\tau \in [0,\tau_m)$
we always have $M_{bst}(\tau)\leq M_{bst}(\tau_m)$. Now we prove it by contradiction. Suppose there exists a ${\tau}' \in [ 0,\tau_m)$ 
such that $M_{bst}({\tau}')\geq M_{bst}(\tau_m)+1$. It is easy to see that 
\begin{align*}
G(M_{bst}(\tau_m)+1,n,\tau_m)<\sqrt{1-\varepsilon_s},
\end{align*}
since $M_{bst}(\tau_m)$ is the largest $m$ satisfying $G(m,n,\tau_m)\geq \sqrt{1-\varepsilon_s}$. By Lemma $6$, it can be observed 
that $G(m,n,\tau)$ increases with $\tau$, whereas decreases with $m$. Thus, we have
\begin{align*}
G(M_{bst}(\tau_m)+1,n,{\tau}')<G(M_{bst}(\tau_m)+1,n,\tau_m)<\sqrt{1-\varepsilon_s} 
\end{align*}
and
\begin{align*}
G(M_{bst}(\tau_m)+1,n,{\tau}')\geq G(M_{bst}({\tau}'),n,{\tau}')\geq \sqrt{1-\varepsilon_s}. 
\end{align*}
We can see that the above two inequalities are contradictory. Thus, for any $\tau \in [0,\tau_m)$
we always have $M_{bst}(\tau)\leq M_{bst}(\tau_m)$ (i.e., $\tau_{bst}^{b}=\tau_m$) and thus 
the eavesdropper-tolerance capability is ${m}^*_{bst}=M_{bst}(\tau_{bst}^{b})$.
\end{IEEEproof}

\subsection{Eavesdropper-Tolerance Capability for Random Relay Selection}\label{sec_4_2}

Applying the same approach, we can establish the following lemma regarding the eavesdropper-tolerance capability for random relay selection.

\emph{Lemma} 7: Consider the network scenario in Fig.\ref{fig_model} with random relay selection scheme. The eavesdropper-tolerance
capability under the security constraint $\varepsilon_s$ and reliability constraint $\varepsilon_t$ is 
\begin{IEEEeqnarray}{rCl}
{m}^*_{ran}= max\{m: G(m,n,\tau_{ran}^b)\geq \sqrt{1-\varepsilon_s}\},
\label{eq_theo2}
\end{IEEEeqnarray}
where 
\begin{align*}
G(m,n,\tau_{ran}^b)=\sum_{k=1}^{m}\binom{m}{k}(-1)^k\Big[(1-e^{-\tau_{ran}^b})c^{k}+e^{-\tau_{ran}^b}\Big]^{n-1},
\end{align*}
$c=\frac{1}{1+\gamma_e}$ and $\tau_{ran}^b$ is the solution of 
\begin{align*}
e^{-\tau}+\frac{1-e^{-(1+\gamma)\tau}}{1+\gamma}=(1-\varepsilon_t)^{\frac{1}{2n-2}}.
\end{align*}

\emph{Remark} 3: Although the exact expressions for ${m}^*_{bst}$ and ${m}^*_{ran}$ are not available, it is easy to calculate them numerically
due to the monotonicity of $G(m,n,\tau)$ with respect to $m$, after calculating the corresponding optimal noise-generating threshold $\tau$ for
these two relay selection schemes.

\section{Numerical Results and Discussion}\label{sec_5}

In this section, we first verify our theoretical model for TOP and SOP through extensive simulations.
We then explore how the number of relays $n$, the \emph{SIR} thresholds $\gamma$, $\gamma_e$, the security constraint $\varepsilon_s$ and the reliability
constraint $\varepsilon_t$ affect the eavesdropper-tolerance capability for opportunistic relaying scheme. Besides, we illustrate the inherent tradeoffs 
between the eavesdropper-tolerance capability and security/reliability constraint. Finally, we compare the opportunistic relaying scheme with
the random relay selection scheme with respect to the eavesdropper-tolerance capability. 

\subsection{Model Validation}\label{sec_5_1}

A simulator was developed in C++ to simulate the message transmission from the source $S$ to the destination $D$ based on the transmission 
scheme in section \ref{sec_2_1}, which is now available at \cite{Mdval2}. The \emph{SIR} threshold for legitimate receivers is fixed as $\gamma=10$ and that
for eavesdroppers is fixed as $\gamma_e = 0.5$. The total number of end-to-end transmissions is fixed as $100000$. The channel varies randomly
and independently from one transmission to another. The simulated TOP (SOP) is calculated as
the ratio of the number of transmissions suffering from transmission outage (secrecy outage) to the total number $100000$. Notice that the 
simulations with other settings can be easily performed by our simulator as well.

\begin{figure}[!t]
\centering
\includegraphics[width=3.5 in]{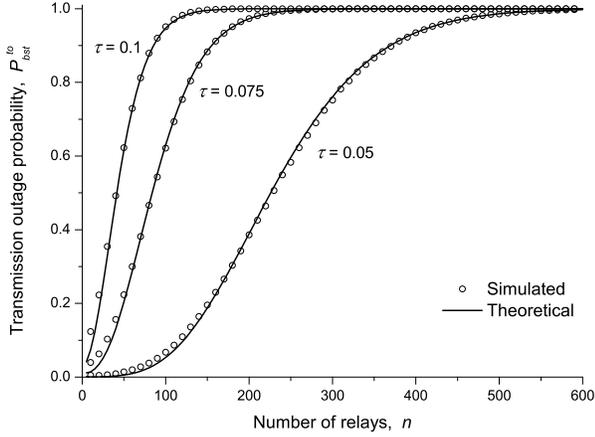}
\caption{TOP for opportunistic relaying $P_{bst}^{to}$ vs. number of relays $n$ with different settings of $\tau$, when $\gamma = 10$.}
\label{fig_TProVal}
\end{figure}

\begin{figure}[!t]
\centering
\includegraphics[width=3.5 in]{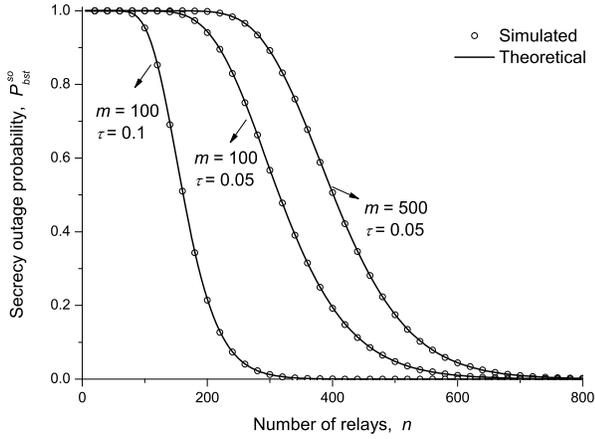}
\caption{SOP for opportunistic relaying $P_{bst}^{so}$ vs. number of relays $n$ with different settings of $m$ and $\tau$, when $\gamma_e = 0.5$.}
\label{fig_SProVal}
\end{figure}

Extensive simulations have been conducted to verify our TOP and SOP models. For the TOP, we considered three
different network scenarios of $\tau=0.05$, $0.075$ and $0.1$, which correspond to low interference, moderate interference and high interference 
compared to the considered network size. For the SOP, we also considered three different network scenarios of $(m=100, \tau=0.05)$,
$(m=100, \tau=0.1)$ and $(m=500,\tau=0.05)$, which correspond to sparse eavesdroppers with low interference, sparse eavesdroppers with high 
interference, and dense eavesdroppers with low interference. The corresponding simulated results and theoretical results are summarized in Fig.
\ref{fig_TProVal} and Fig.\ref{fig_SProVal}.


Fig.\ref{fig_TProVal} and Fig.\ref{fig_SProVal} indicate clearly that the simulated results match nicely with the theoretical ones for both 
TOP and SOP, so our theoretical model can be used to efficiently explore the 
eavesdropper-tolerance capability. A further careful observation of Fig.\ref{fig_TProVal} shows that there is still a very small gap between 
the simulated results and the theoretical results when the number of relays $n$ is very small. For example, for the case that $\tau = 0.075$, 
the simulated value for $P_{bst}^{to}$ is $0.10314$ while the theoretical value is $0.07329$ for $n=30$, compared to the  simulated value of 
$0.46626$ and theoretical value of $0.46645$ for $n=80$. This is because that the Central Limit Theorem  used in deriving our theoretical 
result fails to model the pdf of the total interference $I_1$ and $I_2$ very well for small values of $n$.  We can see from Fig.\ref{fig_TProVal} 
that $P_{bst}^{to}$ increases with $n$. This suggests that although the best relay selection scheme can benefit the transmission as $n$ increases, 
the interferences from the noise-generating relays dominate the tendency of the received $SIR$ at legitimate receivers. By comparing these 
three curves in Fig.\ref{fig_TProVal}, it can also be observed that $P_{bst}^{to}$ increases as $\tau$ increases, which agrees with Lemma $5$. 
This is due to the reason that more interferences will be generated at the intended receiver for lager $\tau$, and thus it is more difficult 
for the receivers to successfully recover the messages. 

We can see from Fig.\ref{fig_SProVal} that $P_{bst}^{so}$ decreases as $n$ increases. This is because more interferences can be generated at 
the eavesdroppers by distributing more relays for a specific $\tau$. By comparing these three curves in Fig.\ref{fig_SProVal}, it can also be 
observed that $P_{bst}^{so}$ increases as $m$ increases while decreases as $\tau$ increases, which agree with Lemma $6$. This is intuitive since 
distributing more eavesdroppers by the adversary would post more potential threats to the end-to-end transmission and increasing $\tau$
would  generate more interferences at the eavesdroppers, so it is more difficult for them to successfully decode the messages. 
 
\subsection{Eavesdropper-tolerance Performances}\label{sec_5_2}

\begin{figure}[!t]
\centering
\includegraphics[width=3.5 in]{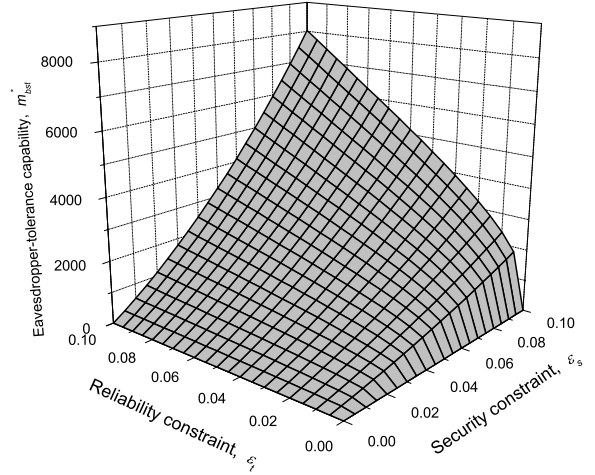}
\caption{Eavesdropper-tolerance capability ${m}^*_{bst}$ for opportunistic relaying vs. reliability constraint $\varepsilon_t$ and security constraint $\varepsilon_s$ with $n = 2000$, $\gamma=10$ and $\gamma_e = 0.5$.}
\label{fig_et_es_m}
\end{figure}

Based on the SOP and TOP models, we now explore the performance of eavesdropper-tolerance capability for opportunistic relaying scheme.
To illustrate the impact of security and reliability constraints on the eavesdropper-tolerance capability, we show in Fig.\ref{fig_et_es_m} 
the behavior of ${m}^*_{bst}$ vs. $\varepsilon_t$ and $\varepsilon_s$ for the network scenario of $n = 2000,\gamma=10,\gamma_e=0.5$, which implies
that the eavesdroppers have a much better decoding ability than the legitimate receivers. We can observe from Fig.\ref{fig_et_es_m} that ${m}^*_{bst}$
increases as $\varepsilon_t$ and $\varepsilon_s$ increase. This reflects that the network can tolerate more eavesdroppers by
relaxing either the security or reliability requirement. A careful observation of Fig.\ref{fig_et_es_m} indicates that $\varepsilon_t$ increases
as $\varepsilon_s$ decreases in order to guarantee a certain level of eavesdropper-tolerance capability. For example, $\varepsilon_t$ has to 
increase from $0.04$ to $0.085$ as $\varepsilon_s$ decreases from $0.03$ to $0.02$ for achieving an eavesdropper-tolerance capacity of about
$1000$. This suggests that either the security or reliability requirement has to sacrifice for the other one in order to achieve a certain 
eavesdropper-tolerance capability. From the above discussions, we can see that there exists clear tradeoffs between the eavesdropper-tolerance
capability and the reliability/security constraint.

\begin{figure}[!t]
\centering
\includegraphics[width=3.5 in]{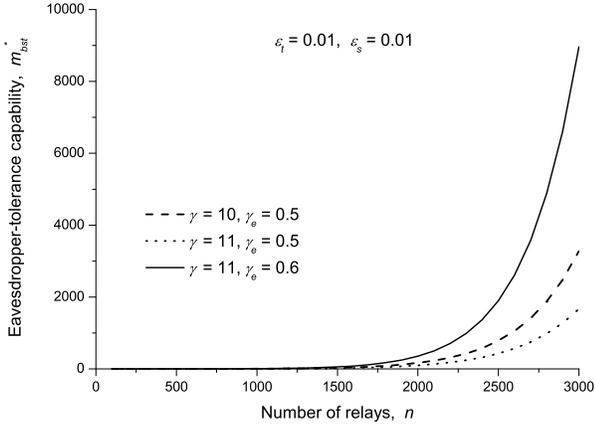}
\caption{Eavesdropper-tolerance capability ${m}^*_{bst}$ for opportunistic relaying vs. number of relays $n$  with $\varepsilon_t = 0.01$, 
and $\varepsilon_s = 0.01$.}
\label{fig_n_m_bst}
\end{figure}

To explore how the number of relays affects the eavesdropper-tolerance capability, we illustrate ${m}^*_{bst}$ vs. $n$ in Fig.\ref{fig_n_m_bst}
with $\varepsilon_t = 0.01$ and $\varepsilon_s = 0.01$ for different settings of $\gamma$ and $\gamma_e$. It can be observed from Fig.\ref{fig_n_m_bst} 
that ${m}^*_{bst}$ increases as $n$ increases. This is because that although the optimal threshold $\tau_{bst}^{b}$ decreases as $n$ increase 
for a specific reliability constraint $\varepsilon_t$, the corresponding expected number of noise-generating nodes increases, so more interferences 
can be generated to suppress the eavesdroppers while the desired reliability can still be ensured. By comparing the three curves, we can also observe 
that ${m}^*_{bst}$ increases as $\gamma_e$ increases, while decreases as $\gamma$ increases. This is intuitive since decreasing the decoding ability 
(i.e.,increasing $\gamma_e$) of the eavesdroppers would decrease the SOP, while decreasing the decoding ability (i.e.,increasing $\gamma$) of legitimate receivers 
would increase the TOP. It is interesting to notice that ${m}^*_{bst}$ increases dramatically when $n$ is above some threshold in Fig.\ref{fig_n_m_bst}. 
For example, for the case that $\gamma=11$ and $\gamma_e=0.6$ this threshold is about $2500$. Thus, distributing more relays would be an efficient 
approach to enhance the eavesdropper-tolerance capability in the construction of a network.

\begin{figure}[!t]
\centering
\includegraphics[width=3.5 in]{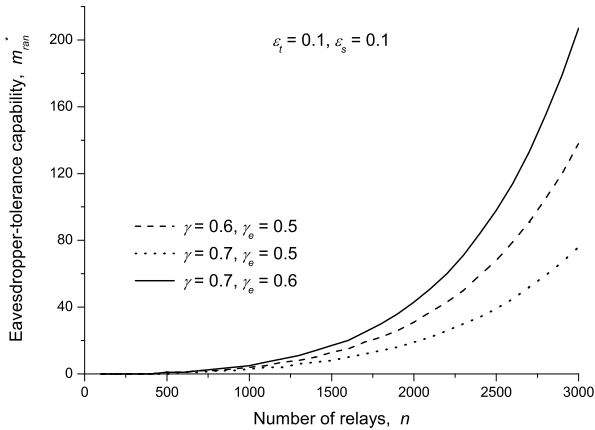}
\caption{Eavesdropper-tolerance capability ${m}^*_{ran}$ for random relay selection vs. number of relays $n$ with $\varepsilon_t = 0.1$, 
and $\varepsilon_s = 0.1$.}
\label{fig_n_m_ran}
\end{figure}

In order to explore the efficiency of the opportunistic relaying scheme, we also illustrate the eavesdropper-tolerance capability ${m}^*_{ran}$
of random relay selection scheme vs. the number of relays $n$ in Fig.\ref{fig_n_m_ran} with $\varepsilon_t = 0.1$ and $\varepsilon_s = 0.1$
for different setting of $\gamma$ and $\gamma_e$. Notice that the security and reliability requirements here are much more relaxed and 
the decoding ability of the legitimate receivers are much more improved than those for opportunistic relaying in Fig.\ref{fig_n_m_bst}. 
For example, $\gamma=0.6$ is much smaller compared to $\gamma=10$ in Fig.\ref{fig_n_m_bst}. That means we consider a much more conservative 
scenario for random relay selection scheme and the network can hardly tolerate any eavesdropper if we consider the same scenario as that 
in Fig.\ref{fig_n_m_bst}. It can be observed from Fig.\ref{fig_n_m_ran} that ${m}^*_{ran}$ also increases as $\gamma_e$ increases, while decreases 
as $\gamma$ increases due to the same reason presented in the discussion of Fig.\ref{fig_n_m_bst}. Even for such a conservative scenario, 
we still can observe from Fig.\ref{fig_n_m_ran} and Fig.\ref{fig_n_m_bst} that the eavesdropper tolerance capability of random relay 
selection is orders of magnitude less than that of opportunistic relaying scheme, especially for large values of $n$. For example,  for the 
case that $\gamma=0.7$ and $\gamma_e=0.6$ in Fig.\ref{fig_n_m_ran} the network can tolerate about $207$ eavesdroppers, which is much less 
than $8959$ eavesdroppers for the case that $\gamma=11$ and $\gamma_e=0.6$ in Fig.\ref{fig_n_m_bst} when $n=3000$. As the eavesdropper-tolerance
capability for random relay selection scheme decreases with $\gamma$, it will decreases to $0$ if we increases $\gamma$ to $11$. This 
implies that the opportunistic relaying scheme can achieve a significantly better eavesdropper-tolerance capability than random relay selection.

\section{Conclusion}\label{sec_6}

This paper established a theoretical framework to analyze the eavesdropper-tolerance capability of a two-hop wireless network, where cooperative 
jamming and opportunistic relaying techniques are adopted to provide secure and reliable end-to-end transmission against passive and independently-operating 
eavesdroppers of unknown location and channel information. We first apply the Central Limit Theorem to model the TOP and SOP in closed form, based 
on which and also the Stochastic Ordering we then develop the model for eavesdropper-tolerance capability analysis. 
Our results indicate that in general more eavesdroppers can be tolerated in the concerned network if a less stringent requirement on both metrics security and reliability is allowed, but a tradeoff between the requirements on these two metrics does exist to ensure a certain level of eavesdropper-tolerance capability.  The results in this paper also reveal that the opportunistic relaying scheme significantly outperforms the random relay selection scheme in terms of the eavesdropper-tolerance 
capability,  and the scheme can guarantee an acceptable eavesdropper-tolerance capability even when a stringent requirement on security and reliability is imposed.

\appendices
\section{Proof of Lemma 1 and 2} \label{app1}
\textbf{Proof of Lemma 1}: From the transmission protocol and the i.i.d fading assumption,
we can easily see that $I_1$ and $I_2$ are the sum of random variables which are smaller than $\tau$
among $n-1$ i.i.d random variables and thus $I_1$ and $I_2$ are independent and identically distributed. 

Now we  take $I_1$ for example to determine the distribution of the total interference in both 
hops. Fist, we define a function 
\begin{align*}
U(x)=\mathbf 1_{x<\tau}(x)\cdot x,
\end{align*}
where 
\begin{align*}
\mathbf 1_{x<\tau}(x)=\left\{\begin{matrix}
 1,&x<\tau \\ 
 0,& otherwise 
\end{matrix}\right.
\end{align*} 
is an indicator function and then $I_1$ can be formulated as 
\begin{align*}
I_1=\sum_{j=1,j\neq b}^{n}U(|h_{R_j,R_b}|^2),
\end{align*} 
which is the sum of $n-1$ i.i.d random variables with pdf given by the following
mixed density and mass function
\begin{align*}
f_{U}(u)=\left\{\begin{matrix}
 e^{-\tau}\delta(u)+e^{-u},& 0\leq u \leq \tau \\ 
 0,& otherwise,
\end{matrix}\right.
\end{align*}
where $\delta(x)$ is the Dirac delta function. The mean and variance of the mixed-type random 
variable $U(|h_{R_j,R_b}|^2)$ can be given by
\begin{align*}
\mu_1= 1-(1+\tau)e^{-\tau}
\end{align*}
and 
\begin{align*}
\sigma_1^2= 1-\tau^{2}e^{-\tau}-(1+\tau)^{2}e^{-2\tau}.
\end{align*}

Therefore, the pdf of $I_1$ can be recursively given by the following mixed density and mass function
\begin{align*}
f(x)=\left\{\begin{matrix}
 e^{-(n-1)\tau}\delta(x)+p_{n-1}(x)e^{-x},& 0\leq x \leq (n-1)\tau \\ 
 0,& otherwise
\end{matrix}\right.,
\end{align*} 
where $p_{n-1}(x)$ is a piecewise function and coincides with different polynomial functions of degree at most $n-2$ 
on each interval $(k\tau,(k+1)\tau]$ for $0\leq k \leq n-2$. However, it is quite difficult to determine the function 
$p_{n-1}(x)$, especially for large $n$. Thus, we approximate it by a normal random variable with mean $\mu=(n-1)\mu_1$ 
and variance $\sigma^2=(n-1)\sigma_1^2$, according to the Central Limit Theorem and its pdf can be approximated by 
\begin{align*}
f(x)\approx \hat{f}(x)=\frac{e^{\frac{(x-\mu)^2}{2\sigma^2}}}{\sigma\sqrt{2\pi}}
\end{align*}
where 
\begin{align*}
\mu=(n-1)\Big[1-(1+\tau)e^{-\tau}\Big]
\end{align*}
and 
\begin{align*}
\sigma=\sqrt{(n-1)\Big[1-\tau^{2}e^{-\tau}-(1+\tau)^{2}e^{-2\tau}\Big]}.
\end{align*}

\textbf{Proof of Lemma 2}: Before deriving the probability in Lemma 2, we first define 
the event that relay $R_k,k=1,\cdots,n$ is selected as the message relay by $A_k$ (i.e., $b=k$). Besides, we 
use a new random variable $S_j$ to define $min\{|h_{S,R_j}|^2,|h_{R_j,D}|^2\}$ for each relay $R_j$. 
It is notable that $S_j,j=1,\cdots,n$ is an exponential random variable with mean $\frac{1}{2}$.
Then, we have
\begin{align*}
A_k\overset{\underset{\mathrm{\Delta}}{}}{=}\bigcap_{j=1,j\neq k}^{n}\left(S_j\leq S_k\right).
\end{align*}   

Now, applying the law of total probability, we have 
\begin{IEEEeqnarray}{rCl}
\IEEEeqnarraymulticol{3}{l}{\mathbb P\Big(|h_{S,R_b}|^2 \geq x ,|h_{R_b,D}|^2\geq y\Big)} \label{eq_pf_lm2_1}\\
&=&\sum_{k=1}^{n}\mathbb P\left(|h_{S,R_k}|^2 \geq x ,|h_{R_k,D}|^2\geq y,A_k\right) \nonumber\\
&=&\sum_{k=1}^{n}\mathbb P\Bigg(|h_{S,R_k}|^2 \geq x ,|h_{R_k,D}|^2\geq y,\bigcap_{j=1,j\neq k}^{n}\left(S_j\leq S_k\right)\Bigg) \nonumber \\
&\overset{(h)}=&\sum_{k=1}^{n}\int_{0}^{\infty}\mathbb P\Bigg(|h_{S,R_k}|^2 \geq x,|h_{R_k,D}|^2\geq y, \nonumber\\
&&S_k=s,\bigcap_{j=1,j\neq k}^{n}\left(S_j\leq s\right)\Bigg)ds \nonumber \\
&=&\sum_{k=1}^{n}\int_{0}^{\infty}\mathbb P\Big(|h_{S,R_k}|^2 \geq x,|h_{R_k,D}|^2\geq y,S_k=s\Big)\nonumber\\
&&\times\mathbb P\Bigg(\bigcap_{j=1,j\neq k}^{n}\left(S_j\leq s\right)\Bigg)ds \nonumber\\
&=&\sum_{k=1}^{n}\int_{0}^{\infty}\mathbb P\Big(|h_{S,R_k}|^2 \geq x,|h_{R_k,D}|^2\geq y,S_k=s\Big) \nonumber\\
&&\times(1-e^{-2s})^{n-1}ds,\nonumber
\end{IEEEeqnarray}
where $(h)$ integrates over all the values $S_k$ can take. 

When $x\geq y\geq 0$, (\ref{eq_pf_lm2_1}) can be reduced to
\begin{IEEEeqnarray}{rCl} 
\IEEEeqnarraymulticol{3}{l}{\mathbb P\Big(|h_{S,R_b}|^2 \geq x ,|h_{R_b,D}|^2\geq y\Big)\label{eq_pf_lm2_2}}\\
&=&\sum_{k=1}^{n}\Bigg\{\int_{x}^{\infty}\mathbb P\Big(|h_{S,R_k}|^2=s,|h_{R_k,D}|^2\geq s\Big)(1-e^{-2s})^{n-1}ds \nonumber\\
&&+\int_{y}^{x}\mathbb P\Big(|h_{S,R_k}|^2>x,|h_{R_k,D}|^2=s\Big)(1-e^{-2s})^{n-1}ds \nonumber\\
&&+\int_{x}^{\infty}\mathbb P\Big(|h_{S,R_k}|^2>s,|h_{R_k,D}|^2=s\Big)(1-e^{-2s})^{n-1}ds\Bigg\} \nonumber\\
&=&2n\int_{x}^{\infty}\frac{(1-e^{-2s})^{n-1}}{e^{2s}}ds+ne^{-x}\int_{y}^{x}\frac{(1-e^{-2s})^{n-1}}{e^s}ds \nonumber\\
&=&1-(1-e^{-2x})^{n-1}+ne^{-x}\int_{e^{-x}}^{e^{-y}}(1-t^2)^{n-1}dt \nonumber\\
&=&1-(1-e^{-2x})^{n-1}+ne^{-x}\Big[\varphi(n,y)-\varphi(n,x)\Big],\nonumber
\end{IEEEeqnarray}
where
\begin{align*}
\varphi(n,x)= e^{-x}{}_2F_1\left(\frac{1}{2},1-n;\frac{3}{2};e^{-2x}\right)
\end{align*}
and ${}_2F_1$ is the Gaussian hypergeometric function.

Similarly, when $0\leq x<y$, (\ref{eq_pf_lm2_1}) can be reduced to 
\begin{IEEEeqnarray}{rCl}
\IEEEeqnarraymulticol{3}{l}{P\Big(|h_{S,R_b}|^2 \geq x ,|h_{R_b,D}|^2\geq y\Big)\label{eq_pf_lm2_3}}\\
&=&1-(1-e^{-2y})^{n-1}+ne^{-y}\Big[\varphi(n,x)-\varphi(n,y)\Big] \nonumber
\end{IEEEeqnarray}
Combining (\ref{eq_pf_lm2_2}) and (\ref{eq_pf_lm2_3}), Lemma 2 then follows.

\bibliographystyle{IEEEtran}
\bibliography{myIEEEref}


%\vfill


\end{document}